\documentclass[a4paper,10pt]{article}
\usepackage[utf8]{inputenc}
\usepackage{amsthm}
\usepackage{amsfonts}
\usepackage{amssymb}	
\usepackage{amsmath}
\usepackage{bbold}
\allowdisplaybreaks
\usepackage{mathtools}
\usepackage[english]{babel}
\usepackage{color}
\usepackage{slashed}
\usepackage{enumerate}
\usepackage{graphicx}
\usepackage{systeme}
\usepackage{dsfont}
\usepackage{relsize}
\usepackage[margin=0.90in]{geometry}
\usepackage{float}
\usepackage{tikz}
\usepackage[labelformat=simple]{subcaption}

\usepackage{graphicx}
\usepackage{bmpsize}
\usepackage{epstopdf}
\usepackage{bbm}
\usepackage{xcolor,etoolbox}
\usepackage{titling}
\thanksmarkseries{arabic} 
\usepackage{authblk}
\newcommand*\samethanks[1][\value{footnote}]{\footnotemark[#1]}
\usepackage{blindtext,graphicx}
\usepackage[absolute]{textpos}
\usepackage{physics}
\usepackage[hang,flushmargin]{footmisc}

\usepackage{xfrac}
\usepackage{nicefrac}
\usepackage{esvect}
\usepackage{cases}
\usepackage{empheq}
\usepackage{stmaryrd}
\usepackage{cancel}
\usepackage[linguistics]{forest}
\usepackage{url}
\usepackage[font=small]{caption}
\usepackage{multirow}
\newcommand{\comsol}{\textit{Comsol Multiphysics}\textsuperscript{\textregistered}}

\usepackage[section]{placeins}
\usepackage[percent]{overpic}
\usepackage{tikz}
\newcommand*\circled[1]{\tikz[baseline=(char.base)]{
\node[shape=circle,draw,inner sep=1pt] (char) {#1};}}
\usepackage{setspace}
\usepackage[giveninits=true,sortcites=true,date=year,maxbibnames=99,doi=false,isbn=false,url=false,eprint=false]{biblatex}
\renewbibmacro{in:}{}
\DeclareFieldFormat{pages}{#1}
\AtEveryBibitem{%
\clearlist{language}%
}
\usepackage{csquotes}

\usepackage{setspace}
\title{Reduced relaxed micromorphic modeling of harmonically loaded metamaterial plates: investigating boundary effects in finite-size structures}
\author{
Plastiras Demetriou,\thanks{Faculty of Architecture and Civil Engineering, TU Dortmund, August-Schmidt-Str. 8, 44227 Dortmund, Germany}
\quad
Gianluca Rizzi\phantom{,}\samethanks[1]
\quad and \quad
Angela Madeo\phantom{,}\thanks{Head of Chair of Structural Mechanics, Faculty of Architecture and Civil Engineering, TU Dortmund, \\ \indent \,\,\,\, August-Schmidt-Str. 8, Dortmund, Germany}
}

\thanksmarkseries{arabic}
\date{\today}
\begin{document}
\maketitle
\begin{abstract}
In this paper, we propose an approach for describing wave propagation in finite-size microstructured metamaterials using a reduced relaxed micromorphic model.
This method introduces an additional kinematic field with respect to the classical Cauchy continua, allowing to capture the effects of the underlying microstructure with a homogeneous model.
We show that the reduced relaxed micromorphic model is not only effective for studying infinite-size metamaterials, but also efficient for numerical simulations and analysis on specimens of finite size.
This makes it an essential tool for designing and optimising metamaterials structures with specific wave propagation properties.
The proposed model's efficiency is assessed through numerical simulations for finite-size benchmark problems, and shows a good agreement for a wide range of frequencies.
The possibility of producing the same macroscopic metamaterial with different but equivalent unit cell ``cuts'' is also analysed, showing that, even close to the boundary, the reduced relaxed micromorphic model is capable of giving accurate responses for the considered loading and boundary conditions.
\end{abstract}

%
%
%
%
%
\section{\label{sec:intro} Introduction}
In recent years, acoustic metamaterials have gathered significant attention for their ability to manipulate mechanical waves in ways that surpass the capabilities of classical materials.
These materials are engineered with ad-hoc microstructures that can effectively control the propagation of mechanical waves, resulting in unique properties such as negative Poisson’s ratio \cite{lakes1987foam,mousanezhad2015hierarchical,d20183d,}, chiral effects \cite{movchan2022waves,garau2018interfacial,frenzel2017three,rizzi2019identification1,rizzi2019identification2,carta2019wave}, band-gaps \cite{liu2000locally,wang2014harnessing,bilal2018architected,celli2019bandgap,el2018discrete,koutsianitis2019conventional,goh2019inverse,zhu2015study,alberdi2021exploring,krodel2015wide,he2018complete,ghavanloo2023formation}, cloaking \cite{buckmann2015mechanical,misseroni2016cymatics,norris2014active,misseroni2019omnidirectional,rossi2020numerical}, focusing \cite{guenneau2007acoustic,cummer2016controlling}, channeling \cite{kaina2017slow,tallarico2017edge,bordiga2019prestress,wang2018channeled,miniaci2019valley,krushynska2018labyrinthine}, negative refraction \cite{willis2016negative,bordiga2019prestress,zhu2015study,srivastava2016metamaterial,lustig2019anomalous,morini2019negative}, and others \cite{carta2019flexural,frecentese2019dispersion,movchan2022wave,movchan2022wave2,touboul2022enhanced,lakes2023experimental,giorgio2017continuum,zheng2022modeling}.
The potential applications of acoustic metamaterials are broad, including also noise reduction, imaging, and communication.
While progress has been made in designing, fabricating, and modeling such materials, several challenges remain.
One of the main obstacles is to achieve the ability of modeling the desired acoustic properties across a wide frequency range and at a large scale, which is crucial for real-world engineering applications.
In recent papers \cite{aivaliotis2020frequency,demore2022unfolding,ramirez2023multi,voss2023modeling,rizzi2022boundary,rizzi2021exploring,rizzi2022metamaterial,rizzi2022towards} we have shown that the reduced relaxed micromorphic model performs well in describing the response of infinite size metamaterials and also for some simple finite size problems.
The importance of micromorphic models for modeling metamaterials and heterogeneous media has also been acknowledged through the development of homogenization techniques in quasi-static regime \cite{alavi2021construction,alavi2022chiral,alavi2022continualization,skatulla2021local,van2020newton,rokovs2020extended,sridhar2020frequency,rokovs2019micromorphic,liu2021computational,
bensoussan2011asymptotic,sanchez1980non,allaire1992homogenization,milton2002theory,hashin1963variational,willis1977bounds,pideri1997second,bouchitte2002homogenization,camar2003determination,suquet1985elements,miehe1999computational,geers2010multi}
as well as, more recently, in the dynamic regime \cite{hill1963elastic,bacigalupo2014second,chen2001dispersive,boutin2014large,craster2010high,andrianov2008higher,hu2017nonlocal,willis2009exact,willis2011effective,willis2012construction,srivastava2014limit,sridhar2018general,srivastava2017evanescent,schwan2021extended}.

However, in order to model dispersion at frequencies higher than the acoustic modes and band-gap, enriched models of the micromorphic type must be used at the homogenized scale.
To our knowledge, the problem proposed in the present paper is the first one which tries to address the fundamental question of homogenized boundary conditions which are representative of complex situations like considering two unit cell ``cuts" for different metamaterials. 
We present this paper as a necessary step to gain the needed insight to proceed towards more complex situations (for example the effect of the cell's ``cut" when the metamaterial is in contact with another homogeneous solid).
\begin{figure}[ht!]
\centering
\begin{overpic}[width=0.85\textwidth]{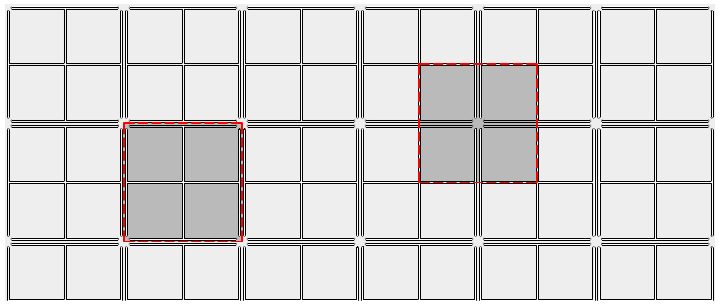}
\put (29.5,20.25) {\makebox[0pt]{\large\circled{$\alpha$}}}
\put (70.5,28.75) {\makebox[0pt]{\circled{$\beta$}}}
\end{overpic}
\caption{
Two of the possible unit cell ``cuts'' that can be chosen and whose periodic tailing builds the same infinite metamaterial.
}
\label{fig:Intro_2_cuts}
\end{figure}

In the present paper, we show for the first time that the reduced relaxed micromorphic model is well adapted to describe the overall behavior of a metamaterials stemming from different unit cell ``cuts", when considering ``free" boundaries and ``thin" Cauchy bars.
The question of studying boundary conditions to be imposed at interfaces between a reduced relaxed micromorphic continuum and another material (e.g. a ``thick" Cauchy bar) to reproduce the response of different ``$\alpha$" and ``$\beta$" cuts is very delicate and will be addressed in forthcoming papers.
The present paper wants to establish that the effect of different cell's cuts does not consistently affect the reduced relaxed micromorphic model's performance as far as simple boundary and loading conditions are considered.

Another issue concerns the choice of the unit cell, the fundamental building block of periodic metamaterials, when dealing with finite-size samples.
The choice of the unit cell ``cut'' (see Fig.~\ref{fig:Intro_2_cuts}) may induce a different response on the metamaterial's boundary that can propagate inside the bulk material.
We show that, for the targeted metamaterial and the chosen applied load, these boundary effects are limited to a region very close to the boundary in almost all cases.
This implies that in this case, possible deviation of the reduced relaxed micromorphic model response may be restricted only to small regions very close to the boundary.
By understanding the properties of metamaterials' unit cells, their associated boundary effects, and by modelling them through the reduced relaxed micromorphic model, we can better understand how to model the propagation of sound waves in finite-size metamaterials' samples for wide ranges of frequencies and different unit cells.
This will allow upscaling and will thus have important implications for fields such as materials science, acoustics, and engineering.
%
%
%
%
%
%
\section{\label{sec:model}The relaxed micromorphic model: a reduced version for dynamics}
We introduce here the equilibrium equations, the associated boundary conditions, and the constitutive relations for a reduced version \cite{ramirez2023multi,demore2022unfolding,rizzi2022metamaterial,rizzi2022boundary,rizzi2022towards,rizzi2021exploring} of the relaxed micromorphic model \cite{neff2014unifying,voss2023modeling,aivaliotis2020frequency,ghiba2015relaxed,madeo2015wave,neff2015relaxed,neff2020identification} for dynamic applications.\footnote{The adjective ``relaxed" was introduced by some of the authors for the specific micromorphic-type continuum model they pioneered some years ago. 
The term ``relaxed" is related to: (\textit{i}) the fact that the curvature term in the strain energy term of the full model is related to the Curl of the micro-distortion $P$ instead than of its entire gradient and (\textit{ii}) contrarily to classical Mindlin-type models, there are no mixed terms of the type $\langle \nabla u-P,\text{sym}\,P\rangle$.}
The equilibrium equations and the boundary conditions are derived with a variational approach thanks to the associated Lagrangian
\begin{equation}
\mathcal{L} (\dot{u},\nabla \dot{u}, \dot{P}, \nabla u, P)
\coloneqq
K (\dot{u},\nabla \dot{u}, \dot{P})
-
W (\nabla u, P)
\ ,
\label{eq:lagrangian}
\end{equation}
where $K$ and $W$ are the kinetic and strain energy, respectively, defined as\footnote{The higher order contributions related to Curl $P$ were neglected in the present paper. 
This choice comes from the observation that, while the curvature term plays a crucial role in the static case (it allows to obtain a specific homogenization formula between the relaxed micromorphic coefficients and the macroscopic stiffnesses when letting $L_{\rm c}\rightarrow0$), its effect is negligible in the dynamic case. However, it can be inferred (see Table~\ref{table:microp} and eq.(\ref{eq:micro_ine_1}))  that the introduced inertia terms also introduce a characteristic length $L_{\rm c}$ which is here of the order of the unit cell size. 
}
\begin{align}
K (\dot{u},\nabla \dot{u}, \dot{P}) \coloneqq&
\dfrac{1}{2}\rho \, \langle \dot{u},\dot{u} \rangle
+ \dfrac{1}{2} \langle \mathbb{J}_{\rm m}  \, \text{sym} \, \dot{P}, \text{sym} \, \dot{P} \rangle
+ \dfrac{1}{2} \langle \mathbb{J}_{\rm c} \, \text{skew} \, \dot{P}, \text{skew} \, \dot{P} \rangle
\label{eq:Ene_kine}
\\*
& 
+ \dfrac{1}{2} \langle \mathbb{T}_{\rm e} \, \text{sym}\nabla \dot{u}, \text{sym}\nabla \dot{u} \rangle
+ \dfrac{1}{2} \langle \mathbb{T}_{\rm c} \, \text{skew}\nabla \dot{u}, \text{skew}\nabla \dot{u} \rangle
\, ,
\notag
\\
W (\nabla u, P) \coloneqq&
\dfrac{1}{2} \langle \mathbb{C}_{\rm e} \, \text{sym}\left(\nabla u -  \, P \right), \text{sym}\left(\nabla u -  \, P \right) \rangle
+ \dfrac{1}{2} \langle \mathbb{C}_{\rm c} \, \text{skew}\left(\nabla u -  \, P \right), \text{skew}\left(\nabla u -  \, P \right) \rangle
\label{eq:Ene_strain}
\\*
& 
+ \dfrac{1}{2} \langle \mathbb{C}_{\rm micro} \, \text{sym}  \, P,\text{sym}  \, P \rangle
\, ,
\notag
\end{align}
where $\langle\cdot , \cdot\rangle$ denote the scalar product, the dot represents a derivative with respect to time, $u \in \mathbb{R}^{3}$ is the macroscopic displacement field, $P \in \mathbb{R}^{3\times 3}$ is the non-symmetric micro-distortion tensor, $\rho$ is the macroscopic apparent density, $\mathbb{J}_{\rm m}$, $\mathbb{J}_{\rm c}$, $\mathbb{T}_{\rm e}$, $\mathbb{T}_{\rm c}$, are 4th order micro-inertia tensors, and $\mathbb{C}_{\rm e}$, $\mathbb{C}_{\rm m}$, $\mathbb{C}_{\rm c}$ are 4th order elasticity tensors (see \cite{rizzi2022boundary,voss2023modeling} for more details).
In particular, we report here the structure of the micro-inertia and the elasticity tensors for the tetragonal class of symmetry and in Voigt notation
\begin{align}
\mathbb{C}_{\rm e}
&=
\begin{pmatrix}
\kappa_{\rm e} + \mu_{\rm e}	& \kappa_{\rm e} - \mu_{\rm e}				& \star & \dots	& 0\\ 
\kappa_{\rm e} - \mu_{\rm e}	& \kappa_{\rm e} + \mu_{\rm e} & \star & \dots & 0\\
\star & \star & \star & \dots & 0\\
\vdots & \vdots	& \vdots & \ddots &\\ 
0 & 0 & 0 & & \mu_{\rm e}^{*}
\end{pmatrix},
\qquad
\mathbb{C}_{\rm micro}
=
\begin{pmatrix}
\kappa_{\rm m} + \mu_{\rm m}	& \kappa_{\rm m} - \mu_{\rm m}				& \star & \dots	& 0\\ 
\kappa_{\rm m} - \mu_{\rm m}	& \kappa_{\rm m} + \mu_{\rm m} & \star & \dots & 0\\
\star & \star & \star & \dots & 0\\
\vdots & \vdots	& \vdots & \ddots &\\ 
0 & 0 & 0 & & \mu_{\rm m}^{*}
\end{pmatrix},
\notag
\\*[5pt]
\mathbb{J}_{\rm m}
&=
\rho L_{\rm c}^2
\begin{pmatrix}
\kappa_\gamma + \gamma_{1} & \kappa_\gamma - \gamma_{1} & \star & \dots & 0\\ 
\kappa_\gamma - \gamma_{1} & \kappa_\gamma + \gamma_{1} & \star & \dots & 0\\ 
\star & \star & \star & \dots & 0\\
\vdots & \vdots & \vdots & \ddots &\\ 
0 & 0 & 0 & & \gamma^{*}_{1}\\ 
\end{pmatrix},
\qquad
\mathbb{T}_{\rm e}
=
\rho L_{\rm c}^2
\begin{pmatrix}
\overline{\kappa}_{\gamma} + \overline{\gamma}_{1} & \overline{\kappa}_{\gamma} - \overline{\gamma}_{1} & \star & \dots	& 0\\ 
\overline{\kappa}_{\gamma} - \overline{\gamma}_{1} &   \overline{\kappa}_{\gamma} + \overline{\gamma}_{1} & \star & \dots & 0\\ 
\star & \star & \star & \dots & 0\\
\vdots & \vdots & \vdots & \ddots &\\ 
0 & 0 & 0 & & \overline{\gamma}^{*}_{1}
\end{pmatrix},
\label{eq:micro_ine_1}
\\*[5pt]
\noalign{\centering
$\mathbb{J}_{\rm c}
=
\rho L_{\rm c}^2
\begin{pmatrix}
\star & 0 & 0\\ 
0 & \star & 0\\ 
0 & 0 & 4\,\gamma_{2}
\end{pmatrix},
\qquad
\mathbb{T}_{\rm c}
=
\rho L_{\rm c}^2
\begin{pmatrix}
\star & 0 & 0\\ 
0 & \star & 0\\ 
0 & 0 & 4\,\overline{\gamma}_{2}
\end{pmatrix} \, ,
\qquad
\mathbb{C}_{\rm c}
= 
\begin{pmatrix}
\star & 0 & 0\\ 
0 & \star & 0\\ 
0 & 0 & 4\,\mu_{\rm c}
\end{pmatrix} \, .
$\\
}
\notag
\end{align}
Only the in-plane components are reported since these are the only ones that play a role in the plane-strain simulations presented in the following sections.
The choice of this particular class of symmetry will be justified in the next section by the choice of the unit cell.
The action functional $\mathcal{A}$ is thus defined as
\begin{align}
\mathcal{A}=\iint\limits_{\Omega \times \left[0,T\right]} 
\mathcal{L} \left(\dot{u},\nabla \dot{u}, \dot{P}, \nabla u, P,\right)
\, \text{d}x \, \text{d}t \, ,
\label{eq:action_func}
\end{align}
and its first variation $\delta \mathcal{A}$ is taken with respect to the kinematic fields $(u,P)$.
Furthermore, it follows from the least-action principle that $\delta \mathcal{A}=0$ uniquely defines both the equilibrium equations and the boundary conditions (both Neumann and Dirichlet).
Thus, the reduced relaxed micromorphic equilibrium equations in strong form are
\begin{align}
\rho\,\ddot{u} - \text{Div}\widehat{\sigma} = \text{Div}\widetilde{\sigma}
\,,
&&
\overline{\sigma} = \widetilde{\sigma} - s
\, ,
\label{eq:equi_equa_RRMM}
\end{align}
where
\begin{align}
\widetilde{\sigma}
&
\coloneqq
\mathbb{C}_{\rm e}\,\text{sym}(\nabla u-P) + \mathbb{C}_{\rm c}\,\text{skew}(\nabla u-P)
\, ,
&
\widehat{\sigma}
&
\coloneqq
\mathbb{T}_{\rm e}\,\text{sym} \, \nabla\ddot{u} + \mathbb{T}_{\rm c}\,\text{skew} \, \nabla\ddot{u}
\,,
\label{eq:equiSigAll}
\\*[5pt]
s
&
\coloneqq
\mathbb{C}_{\rm micro}\, \text{sym} \, P
\, ,
\qquad\qquad\qquad
&
\overline{\sigma}
&
\coloneqq
\mathbb{J}_{\rm m}\,\text{sym} \, \ddot{P} + \mathbb{J}_{\rm c}\,\text{skew} \, \ddot{P}
\, .
\end{align}
The associated homogeneous Neumann boundary conditions are
\begin{align}
\widetilde{t} \coloneqq \left(\widetilde{\sigma} + \widehat{\sigma} \right) \, n = 0 \, ,
\label{eq:tractions}
\end{align}
where $\widetilde{t}$ are the generalized traction and $n$ is the normal to the boundary. 
We also briefly recall here the expression of the traction for a classical \textit{isotropic Cauchy model}
\begin{align}
t \coloneqq \sigma \, n \, ,
&&
\sigma \coloneqq \kappa \, \text{tr}\left(\nabla u\right) \, \mathbb{1} + 2\mu \, \text{dev sym}\nabla u \, ,
\label{eq:tractions_cau}
\end{align}
where $\kappa$ and $\mu$ are the classical bulk and shear moduli, respectively.
%
%
%
\subsection{Identification of the enriched model parameters via dispersion curves fitting}
\begin{figure}[htpb]
\centering
\begin{overpic}[width=0.85\textwidth]{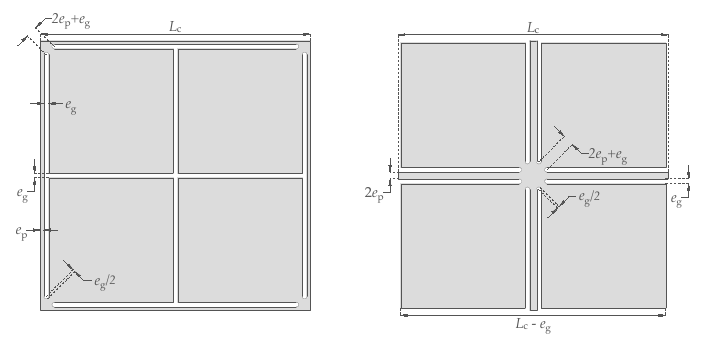}
\put (35,35) {\makebox[0pt]{\large\circled{$\alpha$}}}
\put (85,35) {\makebox[0pt]{\circled{$\beta$}}}
\end{overpic}
\\*
\renewcommand{\arraystretch}{1.3}
\centering
\begin{tabular}{cccccc}
\hline
$L_{\rm c}$ [m]& $e_{\rm g}$ [mm] & $e_{\rm p}$ [mm]& $\rho_{\rm Ti}$  [kg/m$^3$] & $\kappa_{\rm Ti}$ [Pa] & $\mu_{\rm Ti}$ [Pa]
\\
\hline
0.02 & 0.35 & 0.25 & 4400 & 116.7$\times$10$^9$ & 41.8$\times$10$^9$
\\
\hline
\end{tabular}
\caption{
(\textit{top left}) unit cell $\alpha$,  (\textit{top right}) unit cell $\beta$. The two unit cells are equivalent in the sense that they give rise to the same infinite microstructured material, while giving rise to two finite size samples with different geometry on the boundary;
(\textit{bottom}) material and geometrical properties: the size of the unit cell is $L_{\rm c}$, the density $\rho_{\rm Ti}$, the bulk modulus $\kappa_{\rm Ti}$ and the shear modulus $\mu_{\rm Ti}$.
}
\label{fig:unit_cell}
\end{figure}
\begin{figure}[htpb]
\centering
\includegraphics[width=\textwidth]{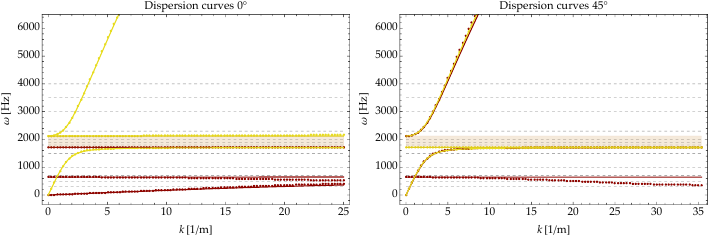}
\caption{
Dispersion curves for $0^\circ$(\textit{left}),  and for $45^\circ$(\textit{right}).
The dots correspond to the solution of the Bloch-Floquet analysis performed on any of the two unit cells in Fig.~\ref{fig:unit_cell} by using \comsol,  while the solid lines represent the analytical expression of the dispersion curves for the reduced relaxed micromorphic model. 
The dashed lines represent the frequencies used in the numerical simulations.
}
\label{fig:Collet_DC}
\end{figure}

In this section we briefly present the reduced relaxed micromorphic parameters identification procedure that is done by the means of fitting the dispersion curves.
On one hand, the dispersion curves (Fig.~\ref{fig:Collet_DC}) of the microstructured material are obtained with a classical Bloch-Floquet analysis performed on any unit cell of the two in Fig.~\ref{fig:unit_cell} by using \comsol.\footnote{The choice of this unit cell is related to the fact that it is the most recent one for which a successful fitting of the RRMM was achieved. 
Indeed, the RRMM has proven to be successful to describe the response of different tetragonal unit cells and it is not limited to this special case. 
However, since it is always necessary a preliminary work to perform the reduced relaxed micromorphic model fitting and since we are here interested to study the effect of different cuts in view of gathering better understanding on the effect of boundary conditions on the overall metamaterial’s response, we opt for the use of this specific unit cell.}
The two unit cells, which we will name $\alpha$ (\textit{left}) and $\beta$ (\textit{right}), give rise to the same dispersion curves since a Bloch-Floquet analysis employs periodic boundary conditions, thus mimicking an infinite domain and the two unit cells shown in Fig.~\ref{fig:unit_cell} are equivalent in the sense that they give rise to the same infinite microstructured material.
On the other hand, dispersion curves for the reduced relaxed micromorphic model are obtained analytically by finding the non trivial solution of the homogeneous equilibrium equations (\ref{eq:equi_equa_RRMM}) under a plane-wave ansatz (for more details see \cite{voss2023modeling}).
The number of independent parameters in the reduced relaxed micromorphic model is 16: 8 of them can be analytically evaluated or analytically related to the other parameters, while the remaining 8 are obtained with an error minimization procedure so that the dispersion curves issued via the reduced relaxed micromorphic model are the closest possible to those issued via Bloch-Floquet analysis (see Fig.~\ref{fig:Collet_DC}).
The parameters with an analytical expression are\footnote{As proved in previous papers, when considering the full relaxed micromorphic model in the static case, rigorous homogenization formula relating the micromorphic coefficient to the macro Cauchy limit can be derived when letting the static characteristic length tend to zero (see \cite{d2020effective}).
We recall here these formulas relating the micromorphic coefficients to the ones associated with the macroscopic Cauchy limit.}
\begin{align}
\rho
&=
\rho_{\rm Ti}\,\frac{A_{\rm Ti}}{A_{\rm tot}}
\,,
&
\kappa_{\rm e}
&=
\frac{\kappa_{\rm m} \, \kappa_{\rm Macro}}{\kappa_{\rm m}-\kappa_{\rm Macro}}
\,,
&
\mu_{\rm e}
&=
\frac{\mu_{\rm m} \, \mu_{\rm Macro}}{\mu_{\rm m}-\mu_{\rm Macro}}
\,,
\notag
\\*
\mu_{\rm e}^*
&=
\frac{\mu_{\rm m}^* \, \mu_{\rm Macro}^*}{\mu_{\rm m}^*-\mu_{\rm Macro}^*}
\,,
&
\kappa_\gamma
&=
\frac{\kappa_{\rm e}+\kappa_{\rm m}}{\rho\,L_{\rm c}^2\,\omega_{\rm p}^2}
&
\gamma_1
&=
\frac{\mu_{\rm e}+\mu_{\rm m}}{\rho\,L_{\rm c}^2\,\omega_{\rm s}^2}
\,,
\label{eq:analytical_parameters}
\\*
\gamma_1^*
&=
\frac{\mu_{\rm e}^*+\mu_{\rm m}^*}{\rho\,L_{\rm c}^2\,\omega_{\rm ss}^2}
\,,
&
\gamma_2&=\frac{\mu_{\rm c}}{\rho\,L_{\rm c}^2\,\omega_{\rm r}^2}
\,.
\notag
\end{align}
where $A_{\rm Ti}$ and $A_{\rm tot}$ are respectively the area of Titanium and the total area (including the voids) of the unit cell, $\omega_{\rm p}$, $\omega_{\rm r}$, $\omega_{s}$, and $\omega_{ss}$ are the cut-off frequencies, namely the frequencies for a vanishing wavenumber $k=0$, $\kappa_{\rm Macro}$, $\mu_{\rm Macro}$, and $\mu_{\rm Macro}^*$ are the \textit{macro-parameters}, which represent the stiffness of the microstructured material for the long-wavelength limit and can be obtained thanks to the relations of the slope (wave speed) for $k=0$ of the acoustic branches of the dispersion curves (or with classical static test with periodic boundary conditions), while $L_{\rm c}$ is the length of the side of the unit cell, i.e. $L_{\rm c}=0.02$ m.

\begin{table}[htpb]
\renewcommand{\arraystretch}{1.3}
\centering
\begin{tabular}{cccccc}
\hline
$L_{\rm c}$ [m]& $\kappa_{\rm e}$ [Pa]& $\mu_{\rm e}$ [Pa]& $\mu_{\rm e}^{*}$ [Pa]
\\
\hline
0.02 & 2.51$\times$10$^9$ & 2.39$\times$10$^9$ & 1.20$\times$10$^6$
\\
\hline
\hline
$\mu_{\rm c}$ [Pa]& $\kappa_{\rm m}$ [Pa]& $\mu_{\rm m}$ [Pa]& $\mu_{\rm m}^{*}$ [Pa]
\\
\hline
1.11$\times$10$^4$ & 4.54$\times$10$^9$ & 4.43$\times$10$^9$ & 4.18$\times$10$^{10}$
\\
\hline
\hline
$\kappa_{\gamma}$ [-]& $\gamma_1$ [-]& $\gamma^{*}_1$ [-]& $\gamma_2$ [-]
\\
\hline
1016.56 & 983.36 & 9234.89 & 0.02
\\
\hline
\hline
$\overline{\kappa}_{\gamma}$ [-]& $\overline{\gamma}_1$ [-]& $\overline{\gamma}^{*}_1$ [-]& $\overline{\gamma}_2$ [-]
\\
\hline
5.45 & 3.09 & 1.45$\times$10$^{-9}$ & 1.87
\\
\hline
\hline
$\kappa_{\rm Macro}$ [Pa]& $\mu_{\rm Macro}$ [Pa]& $\mu^{*}_{\rm Macro}$ [Pa]& $\rho$ [kg/m$^3$]
\\
\hline
1.62$\times$10$^9$ & 1.55$\times$10$^9$ & 1.20$\times$10$^6$ & 3840.77
\\
\hline
\end{tabular}
\caption{
Values of the elastic parameters, the micro-inertia parameters, the characteristic length $L_{\rm c}$, and the apparent density $\rho$ for the reduced relaxed micromorphic model calibrated on the metamaterial whose building block is any of the two unit cells in Fig.~\ref{fig:unit_cell}. In the last row it is reported the \textit{macro-parameters}, i.e., the corresponding long-wavelength limit Cauchy material coefficients \cite{neff2020identification,rizzi2021exploring}.
}
\label{table:microp}
\end{table}
As already remarked, the remaining 8 parameters $\kappa_{\rm m}$, $\mu_{\rm m}$, $\mu_{\rm m}^*$, $\mu_{\rm c}$, $\overline{\kappa}_\gamma$, $\overline{\gamma}_1$, $\overline{\gamma}_1^*$, and $\overline{\gamma}_2$ are obtained by minimizing the distance between the dispersion curves obtained via Bloch-Floquet analysis and the ones of the equivalent reduced relaxed micromorphic model through a fitting procedure (for more details see \cite{voss2023modeling}).
All the material parameters of the reduced relaxed micromorphic model characterising the microstructured material of Fig.~\ref{fig:unit_cell} are summarised in Table~\ref{table:microp}, and the plots of the two sets of curves are shown in Fig.~\ref{fig:Collet_DC}.\footnote{The dispersion curves depicted in Fig.~\ref{fig:Collet_DC} represent the first 6 modes of an infinite metamaterial constituted by a periodic repetition of the unit cell in Fig. \ref{fig:Collet_DC}: yellow curves are associated to pressure modes, while red curves represent shear modes. 
For a direction of propagation of 45° the acoustic shear and pressure mode are almost superimposed, and the same is for the shear and pressure optic modes starting around 2000 Hz. The band gap region is highlighted with a light brown color. 
Clearly, the full microstructured system may exhibit an infinite number of modes when increasing frequency. On the other hand, the considered micromorphic model can only reproduce the first six of these modes. 
To introduce higher frequency modes  also in the micromorphic modeling framework, extra micro-related degrees of freedom should be introduced with respect to the micro-distortion tensor $P$ alone. 
However, we are interested in this paper to the lower frequency behaviors of the considered metamaterial and, in particular, to the possibility of adequately describing dispersion and band-gaps in this frequency interval.
}
%
%
%
%
%
%
\section{Finite element simulations set-up}
\label{Sec:FE_S}
In this section we present the setting-up of the numerical simulations on a finite-size metamaterial both with a microstructured Cauchy model and the reduced relaxed micromorphic model. 
%
%
%
\subsection{Microstructured materials simulations set-up}
All the 2D simulations presented here have been performed under a plane-strain assumption and with a \textit{time-harmonic} ansatz.
The two microstructured materials presented in this work have been built as a regular grid of finite-size ($16\times 16$ unit cells of side $L_{\rm c}=0.02$ m), whose building blocks are the unit cells made up of Titanium shown in Fig.~\ref{fig:unit_cell}.
The resulting metamaterials are connected to two slender homogeneous Cauchy bars made up of Titanium.
\begin{figure}[htpb]
\centering
\includegraphics[width=\textwidth]{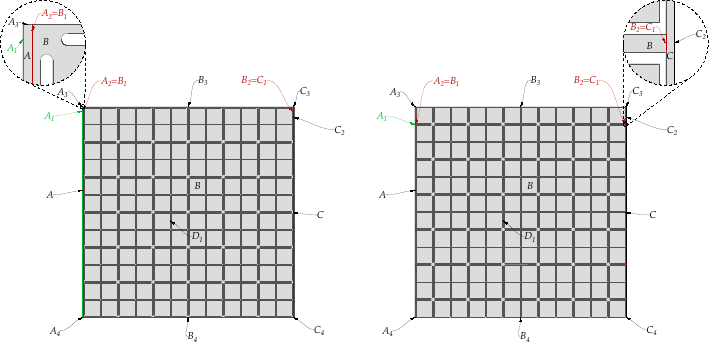}
\caption{
Schematic view of the geometry and the labeling of the boundaries and interfaces for (\textit{left}) the microstructured material built out of the unit cell $\alpha$ and (\textit{right}) the microstructured material built out of the unit cell $\beta$.
}
\label{fig:BC_struc}
\end{figure}
The following boundary and interface conditions have been enforced (see Fig.~\ref{fig:BC_struc})
\begin{align}
\begin{array}{rlllll}
u_1\vert_{A_1}   & \!\!\!\! = \overline{u}      &  \text{and}\quad \,\,\,\, u_2\vert_{A_1} \to \text{free}                    & \quad (\text{prescribed disp. - green})  & \text{on} &  A_1
\\*
u\vert_{A_2}     & \!\!\!\! = u\vert_{B_1}      &    \text{and}\quad \sigma \, n\vert_{A_2} = \sigma \, n\vert_{B_1}   & \quad(\text{perfect contact - red})    & \text{on} &  A_2 \equiv  B_1
\\*
u\vert_{B_2}     & \!\!\!\! = u\vert_{C_1}      &    \text{and}\quad \sigma \, n\vert_{B_2} = \sigma \, n\vert_{C_1}   & \quad(\text{perfect contact - red})    & \text{on} &  B_2 \equiv  C_1
\\*
\sigma \, n      & \!\!\!\! = 0                 &                                                                      & \quad(\text{stress free - black})       & \text{on} &  A_3, A_4, B_3, B_4,
\\*
&                              &                                                                      &                                        &           & C_2, C_3, C_4, D_1
\end{array}
\label{eq:BC_MS}
\end{align}
where the magnitude of the harmonic prescribed displacement $\overline{u}$ is 1\% of the size of the specimen.\\
In particular we set $ \overline{u}=u_0 \, e^{-i\omega t} $ with $u0=3.2$ mm.
The simulations have been performed by using the \textit{Solid Mechanics} physics package of \comsol.
In order to ease the convergence of the analysis, we introduced a small amount of numerical isotropic damping ($\eta=0.002$).
%
%
%
\subsection{Reduced relaxed micromorphic continuum simulations set-up}
\begin{figure}[htpb]
\centering
\includegraphics[width=0.5\textwidth]{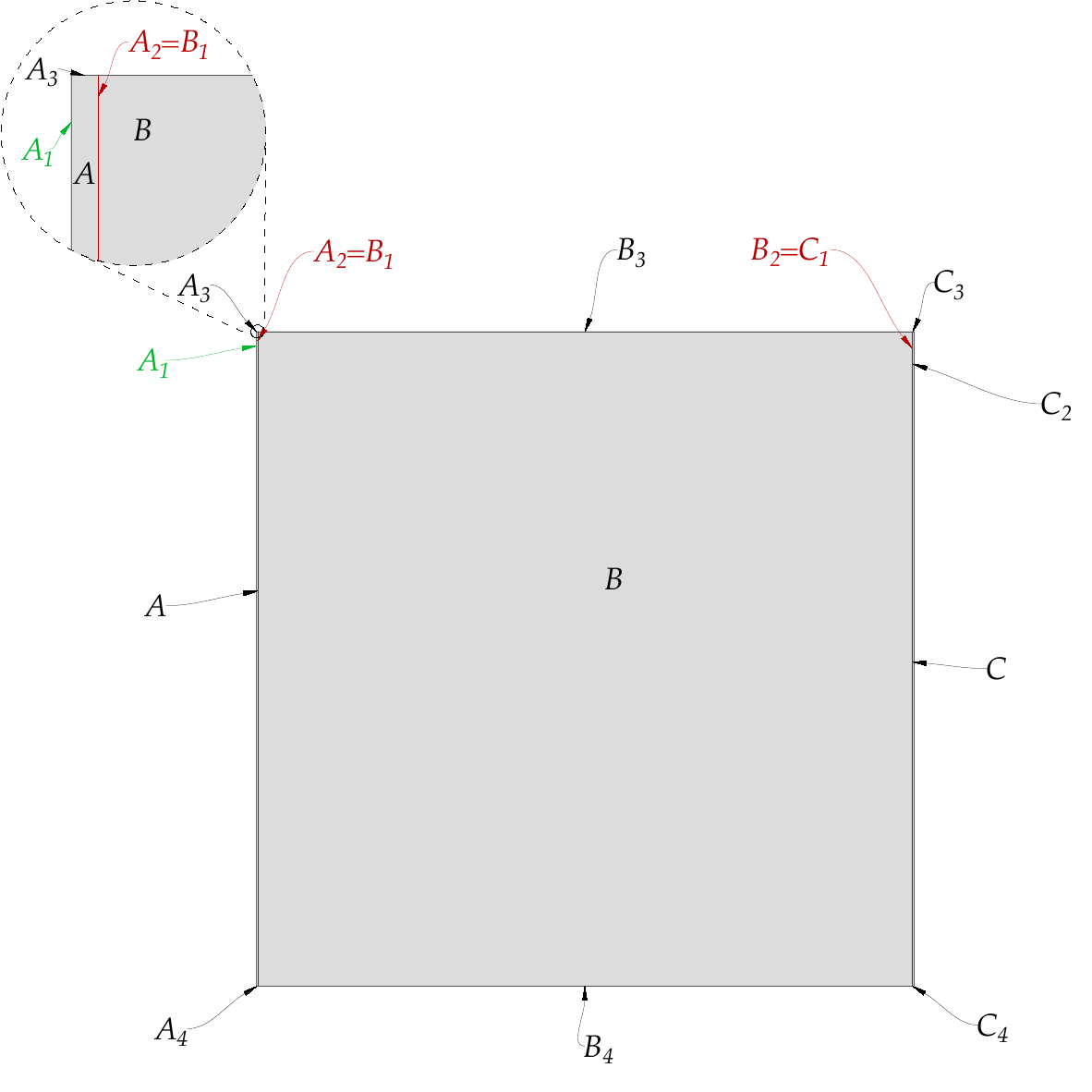}
\caption{
Schematic view of the geometry and the labeling of the boundaries and interfaces for the equivalent reduced relaxed micromorphic material.
}
\label{fig:BC_conti}
\end{figure}

The microstructured material is here modeled with the reduced relaxed micromorphic model, which is characterised by the material parameters in Table~\ref{table:microp}.
In addition, the following boundary and interface conditions have been enforced (see Fig.~\ref{fig:BC_conti})
\begin{align}
\begin{array}{rlllll}
u_1\vert_{A_1}                  & \!\!\!\! = \overline{u}      &    \text{and}\quad \,\,\,\, u_2\vert_{A_1} \to \text{free}                                  & \quad (\text{prescribed disp. - green})     & \text{on} & A_1
\\*
u\vert_{A_2}                    & \!\!\!\! = u\vert_{B_1}      &    \text{and}\quad \sigma \, n\vert_{A_2} = \widetilde{\sigma} \, n\vert_{B_1}     & \quad(\text{perfect contact - red})       & \text{on} & A_2 \equiv  B_1
\\*
u\vert_{B_2}                    & \!\!\!\! = u\vert_{C_1}      &    \text{and}\quad \widetilde{\sigma} \, n\vert_{B_2} = \sigma \, n\vert_{C_1}     & \quad(\text{perfect contact - red})       & \text{on} & B_2 \equiv  C_1
\\*
\sigma \, n                     & \!\!\!\! = 0                 &                                                                                    & \quad\text{(stress free - black)}          & \text{on} & A_3, A_4, C_2, C_3, C_4
\\*
(\widetilde{\sigma} + \widehat{\sigma}) \, n         & \!\!\!\! = 0                 &                                                                                    & \quad\text{(stress free - black)}          & \text{on} &  B_3, B_4
\end{array}
\label{eq:BC_MM}
\end{align}
where again, the magnitude of the prescribed harmonic displacement $\overline{u}$ is 1\% of the size of the specimen.
The effective homogeneous material modeled with the reduced relaxed micromorphic model is also embedded between two slender homogeneous Cauchy bars made up of Titanium.
The simulations have been performed by using the \textit{Weak Form PDE} physics package of \comsol.
This package requires the implementation of the expression of the Lagrangian~(\ref{eq:lagrangian}) and the appropriate boundary and interface conditions.
To have a consistent comparison with the results from the microstructured material, we have introduced the same small amount of numerical isotropic viscous damping ($\eta=0.002$) also in this case.
%
%
%
%
%
%
\section{Results and comparison}
In this section we show the results issued by the numerical simulations described in Section \ref{Sec:FE_S} for the values of frequencies highlighted in Figure \ref{fig:Collet_DC} (dashed lines).

\begin{figure}[htpb]
\centering
\includegraphics[width=\textwidth]{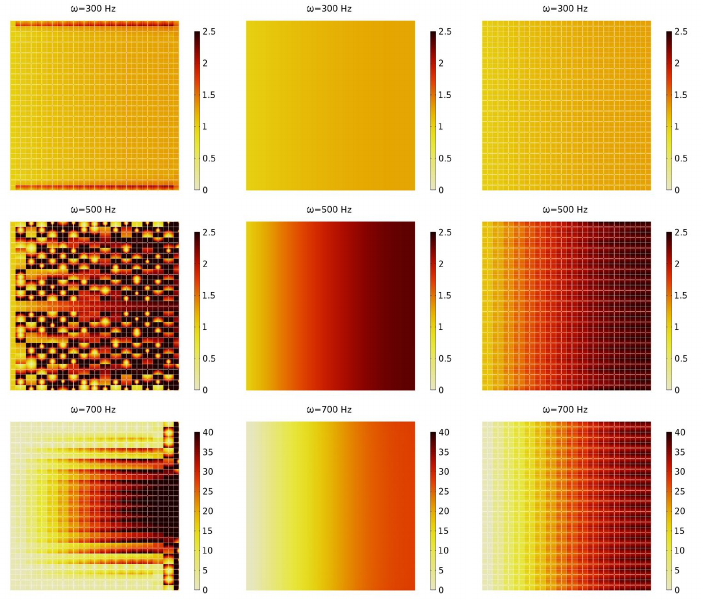}
\caption{
Norm of the displacement field $\lvert u \rvert$ divided by the amplitude of the assigned harmonic displacement $\overline{u}$ for 300, 500, and 700 Hz for (\textit{left}) the microstructured material whose building block is the unit cell $\alpha$, (\textit{center}) the equivalent reduced relaxed micromorphic material, and (\textit{right}) the microstructured material whose building block is the unit cell $\beta$.
}
\label{fig:1}
\end{figure}

\begin{figure}[htpb]
\centering
\includegraphics[width=\textwidth]{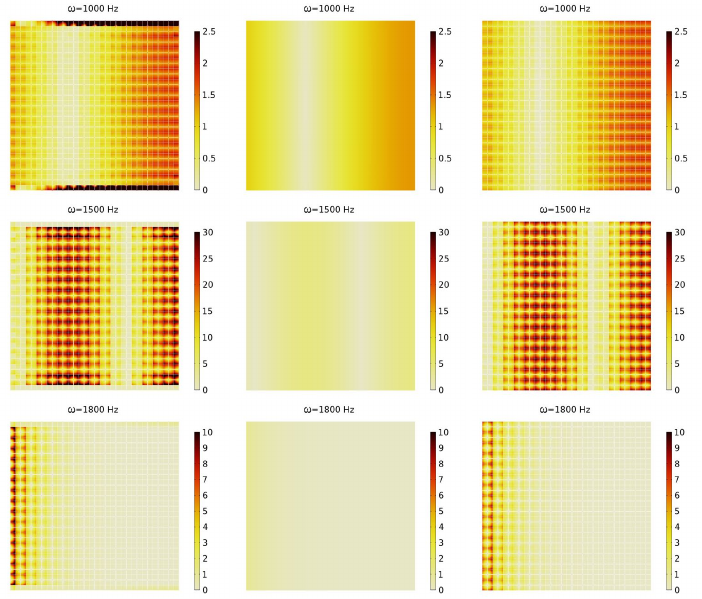}
\caption{
Norm of the displacement field $\lvert u \rvert$ divided by the amplitude of the assigned harmonic displacement $\overline{u}$ for 1000, 1500, and 1800 Hz for (\textit{left}) the microstructured material whose building block is the unit cell $\alpha$, (\textit{center}) the equivalent reduced relaxed micromorphic material, and (\textit{right}) the microstructured material whose building block is the unit cell $\beta$.
}
\label{fig:2}
\end{figure}

\begin{figure}[htpb]
\centering
\includegraphics[width=\textwidth]{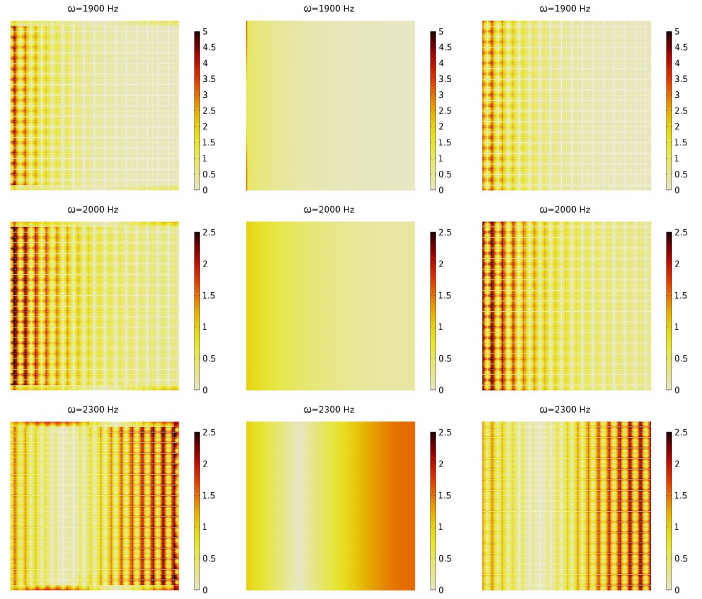}
\caption{
Norm of the displacement field $\lvert u \rvert$ divided by the amplitude of the assigned harmonic displacement $\overline{u}$ for 1900, 2000, and 2300 Hz for (\textit{left}) the microstructured material whose building block is the unit cell $\alpha$, (\textit{center}) the equivalent reduced relaxed micromorphic material, and (\textit{right}) the microstructured material whose building block is the unit cell $\beta$.
}
\label{fig:3}
\end{figure}

\begin{figure}[htpb]
\centering
\includegraphics[width=\textwidth]{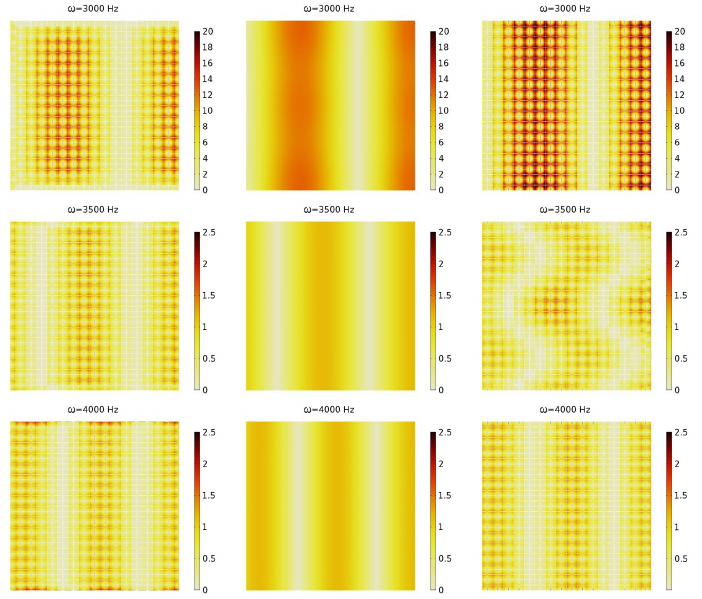}
\caption{
Norm of the displacement field $\lvert u \rvert$ divided by the amplitude of the assigned harmonic displacement $\overline{u}$ for 3000, 3500, and 4000 Hz for (\textit{left}) the microstructured material whose building block is the unit cell $\alpha$, (\textit{center}) the equivalent reduced relaxed micromorphic material, and (\textit{right}) the microstructured material whose building block is the unit cell $\beta$.
}
\label{fig:4}
\end{figure}

In Figs.~(\ref{fig:1})-(\ref{fig:4}) the structural response for different frequencies is given for (\textit{left}) the microstructured material built out of the unit cell $\alpha$, (\textit{center}) the equivalent reduced relaxed micromorphic material, and (\textit{right}) the microstructured material built out of the unit cell $\beta$.
It can be inferred by direct inspection of these figures that the unit cell's cut shown in the \textit{right panel} of Fig.~\ref{fig:unit_cell} gives rise to a macroscopic response which is better captured by the reduced relaxed micromorphic model, except for the frequency $\omega=3500$ Hz at which a resonant mode is predominant (see the \textit{right panel} of the second row Fig.~\ref{fig:4}).
However, we can also notice that, except for some small regions close to the boundary, also the behaviour of the cut shown in the \textit{left panel} of Fig.~\ref{fig:unit_cell} is captured at an acceptable level of agreement.
Exceptions arise for the frequency $\omega=500$ Hz and $\omega=700$ Hz at which, once again, microstructure related resonant modes might become predominant.\footnote{Microstructure related resonant modes are observed by direct inspection of the displacement field inside the unit cell for the microstructured simulation.}
\FloatBarrier
%
%
%
%
%
%
\section{Discussion}
\begin{figure}[htpb]
\centering
\includegraphics[width=0.55\columnwidth]{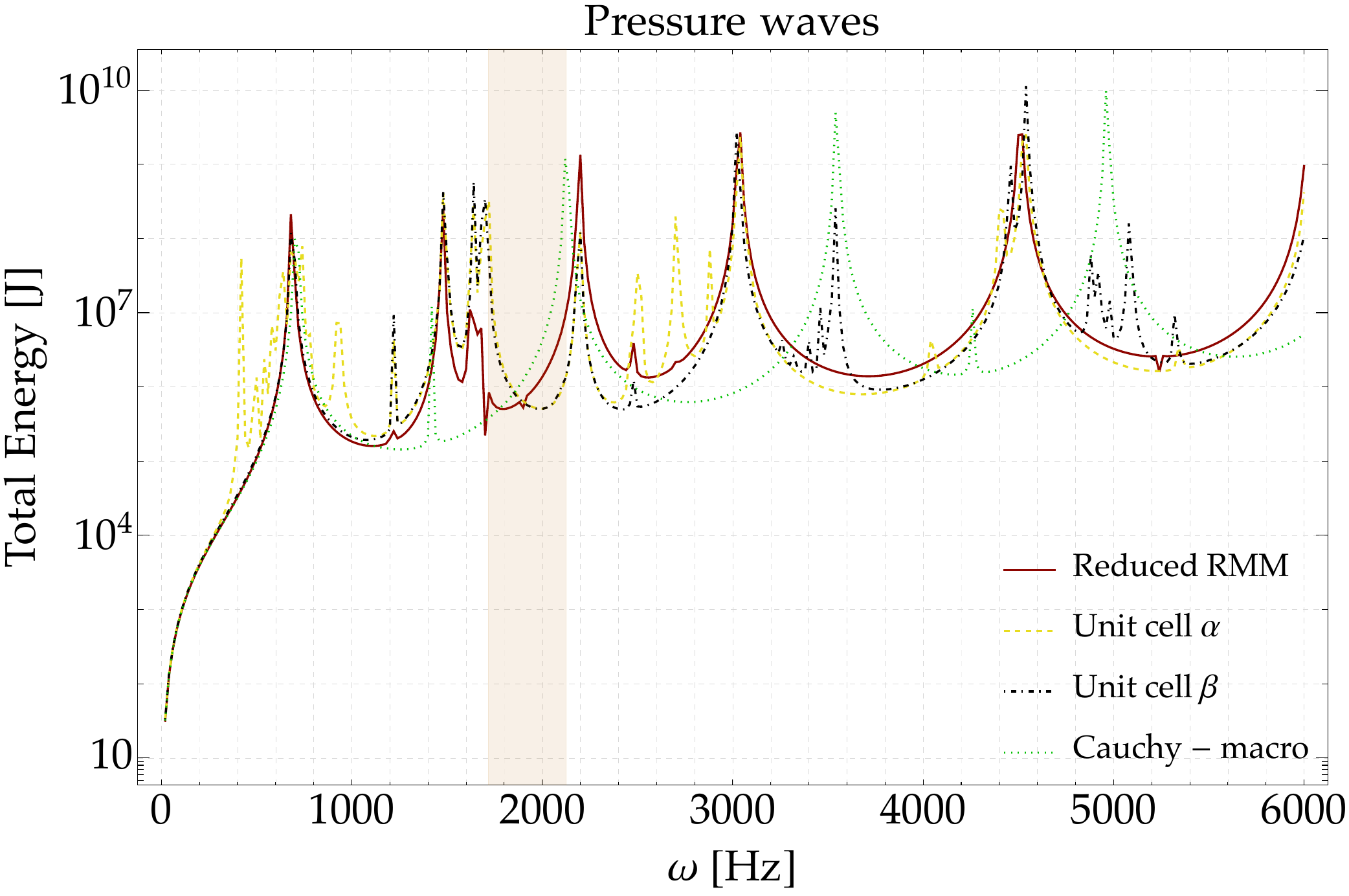}
\caption{
Comparison between total energy of the microstructured material $\alpha$ (yellow dashed), the microstructured material $\beta$ (black dot-dashed), the reduced relaxed micromorphic continuum (solid red), and the equivalent macro Cauchy continuum (green dotted) in the frequency range [0,6000] Hz.
}
\label{fig:energy_plots}
\end{figure}

In Fig.~\ref{fig:energy_plots} we show how the comparison between the total energy of (i) the microstructured material built with the unit cell $\alpha$ (yellow dashed), (ii) the microstructured material built with the unit cell $\beta$ (black dot-dashed), (iii) the reduced relaxed micromorphic continuum (solid red), and  (iv) the equivalent macro Cauchy continuum (green dotted) in the frequency range [0,6000] Hz.

\begin{figure}[htpb]
\centering
\includegraphics[width=0.85\textwidth]{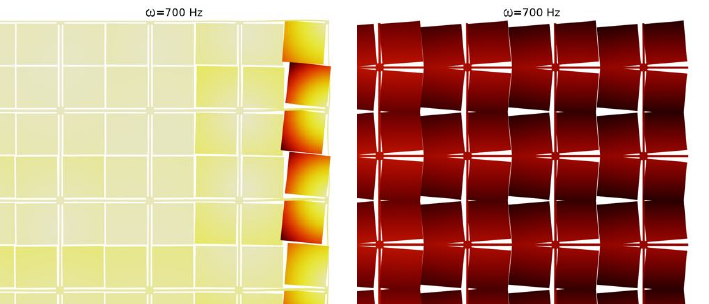}
\caption{
Detail of the deformation for $\omega=700$ Hz for (\textit{left}) the structure based on the unit cell $\alpha$ and for (\textit{right}) the structure based on the unit cell $\beta$.
It can be seen that the unit cell $\alpha$ has a non-symmetric response that propagates along the vertical boundary, while the unit cell $\beta$ gives clearly rise to a symmetric response.
More particularly, in the ``symmetric response" (\textit{right}) the internal resonators rotate of the same quantity at the top and at the bottom, so that the thin beams remain undeformed. In the ``asymmetric response" (\textit{left}) the rotations of top and bottom element do not compensate each other. 
This results in the bending of the thin beams inside the unit cell. 
To make the plot clearer, the homogeneous Cauchy bar at the end of the specimen has been removed from the plot.}
\label{fig:particular_Collet_1_2_f0700}
\end{figure}

\begin{figure}[htpb]
\centering
\includegraphics[width=0.85\textwidth]{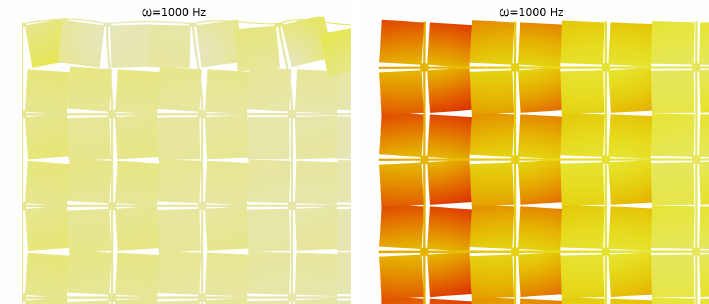}
\caption{
Detail of the deformation for $\omega=1000$ Hz for (\textit{left}) the structure based on the unit cell $\alpha$ and for (\textit{right}) the structure based on the unit cell $\beta$.
It can be seen that the unit cell $\alpha$ has a localised non-symmetric response on the top left corner, while the unit cell $\beta$ gives clearly rise to a symmetric response.
Also in this case the ``non symmetric" response of the resonators implies bending of the thin beams inside the unit cell, while the ``symmetric response" leaves the thin beams undeformed.}
However, the unit cell $\alpha$ recovers a symmetric response while moving away form the boundary.
To make the plot clearer, the homogeneous Cauchy bar at the end of the specimen has been removed from the plot.
\label{fig:particular_Collet_1_2_f1000}
\end{figure}

\begin{figure}[htpb]
\centering
\includegraphics[width=0.85\textwidth]{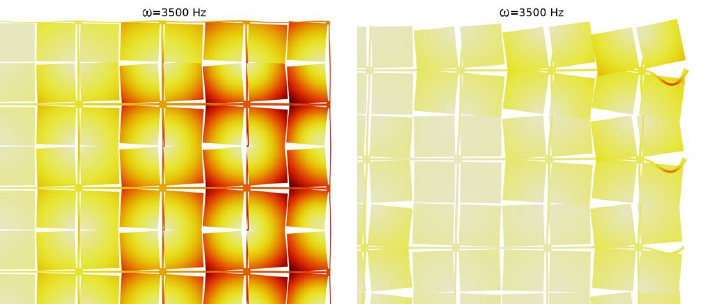}
\caption{
Detail of the deformation for $\omega=3500$ Hz for (\textit{left}) the structure based on the unit cell $\alpha$ and for (\textit{right}) the structure based on the unit cell $\beta$.
It can be seen that the unit cell $\alpha$ has an overall symmetric response on the boundary, while the unit cell $\beta$ gives clearly rise to a prominent non-symmetric one.
Once again the ``non symmetric" response gives rise to pronounced bending of the internal thin beams, while the ``symmetric response" leaves them almost undeformed.
To make the plot clearer, the homogeneous Cauchy bar at the end of the specimen has been removed from the plot.}
\label{fig:particular_Collet_1_2_f3500}
\end{figure}

In Figs.~\ref{fig:particular_Collet_1_2_f0700}-\ref{fig:particular_Collet_1_2_f1000}-\ref{fig:particular_Collet_1_2_f3500} we show a detail of the deformation close to the boundary for the simulations at which a disagreement with the reduced relaxed micromorphic model was detected in one of the two microstructured simulations.
It seems to be the case that in all the simulations where the macroscopic response of the reduced relaxed micromorphic model deviates from the microstructured one, important bending of the structural elements constituting the unit cell occurs (we better describe this bending in the captions of Figures \ref{fig:particular_Collet_1_2_f0700}-\ref{fig:particular_Collet_1_2_f3500}).
\FloatBarrier
%
%
%
\subsection{Effects of the size of the metastructure on the propagation of the boundary localization}
\begin{figure}[htpb]
\centering
\includegraphics[width=0.95\textwidth]{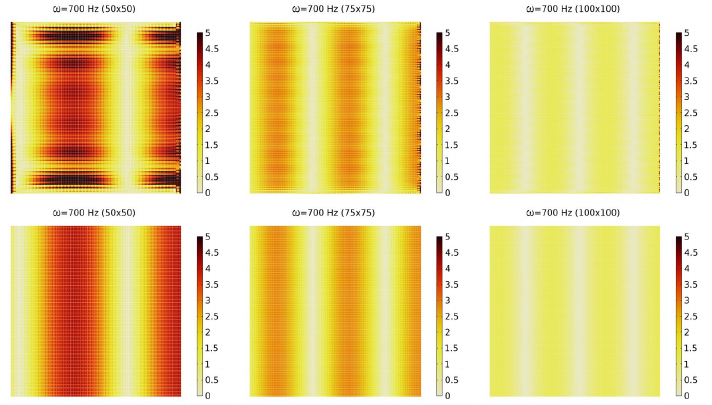}
\caption{
Norm of the displacement field $\lvert u \rvert$ divided by the amplitude of the assigned harmonic displacement $\overline{u}$ for 700 Hz for (\textit{top}) the microstructured material whose building block is the unit cell $\alpha$, (\textit{bottom}) the microstructured material whose building block is the unit cell $\beta$, for a $50\times50$, $75\times75$, and $100\times100$ unit cells metastructures.
The cut $\alpha$ shows an important boundary effect that propagates in the bulk material up to a $50\times50$ unit cells metastructure while it fades form $75\times75$.
The cut $\beta$ does not show noticeable boundary effects regardless the size.
}
\label{fig:Collet_1_2_f0700_50_75_100}
\end{figure}

In order to further investigate how these boundary effects persist while increasing the size of the domain we present here the results for an increasingly big metastructure domain for the frequencies $\omega=500$ Hz and $\omega=700$ Hz.
\begin{figure}[htpb]
\centering
\includegraphics[width=0.95\textwidth]{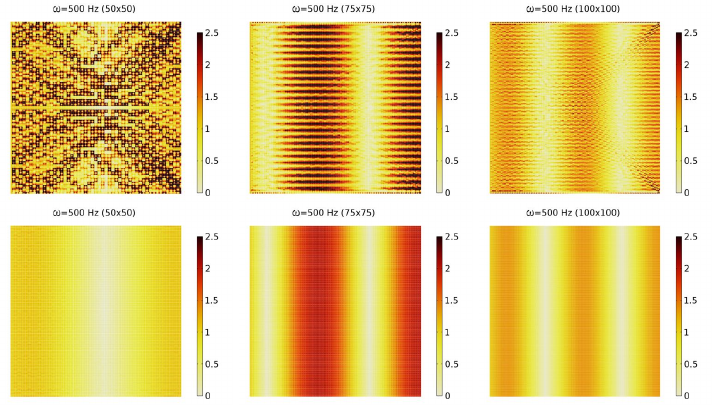}
\caption{
Norm of the displacement field $\lvert u \rvert$ divided by the amplitude of the assigned harmonic displacement $\overline{u}$ for 500 Hz for (\textit{top}) the microstructured material whose building block is the unit cell $\alpha$, (\textit{bottom}) the microstructured material whose building block is the unit cell $\beta$, for a $50\times50$, $75\times75$, and $100\times100$ unit cells metastructures.
The cut $\alpha$ shows an important boundary effect that propagates in the bulk material up to a $100\times100$ unit cells metastructure, while the cut $\beta$ does not show noticeable boundary effects regardless the size.
}
\label{fig:Collet_1_2_f0500_50_75_100}
\end{figure}
In Figs.~\ref{fig:Collet_1_2_f0500_50_75_100}-\ref{fig:Collet_1_2_f0700_50_75_100}
we show how the displacement field changes for a $50\times50$, a $75\times75$, and a $100\times100$ unit cells metastructures: in the top row we report the results for the cut $\alpha$ while in the bottom row the one for the cut $\beta$.
For $\omega=500$ Hz we can see how the boundary effects start to become negligible from a $100\times100$ unit cells metastructure, while before their effect in the bulk is still relevant.
On the contrary, for $\omega=700$ Hz we can see how the boundary effects start to become negligible from a $75\times75$ unit cells metastructure, while in a $100\times100$ their effect is completely relegated to the boundaries and does not affect the bulk material.
\FloatBarrier
%
%
%
%
%
%
\section{Conclusions}
The reduced relaxed micromorphic model has proven to be an effective tool for analysing and predicting the behavior of microstructured materials also at finite scales.
This finding suggests that the enriched model is capable of accurately representing the complex physical interactions and dynamics that occur within these materials, across a broad range of frequencies unless intense small-scale-related resonances are activated.
Such a capability is invaluable for developing new technologies and improving existing ones at the scale of the engineers, where a thorough understanding of the behavior of microstructured materials is essential.
We showed that when considering simple load conditions and leaving the metamaterial's boundary ``free'', the reduced relaxed micromorphic model is effective notwithstanding the choice of unit cell ``cut".
However, the question of how to model the effect of the cell's ``cut'' in more complex situations (e.g. metamaterial in contact with another solid), remains open.
One possible solution could be to enrich the boundary conditions of the reduced relaxed micromorphic model or to explore additional terms introducing specific characteristic lengths.
Addressing this issue is critical for achieving even greater accuracy and reliability in modeling microstructured materials and will be addressed in forthcoming papers.


\vspace{10pt}
\begingroup
\scriptsize
\noindent
\textbf{Acknowledgements.}
Angela Madeo, Plastiras Demetriou, and Gianluca Rizzi acknowledge support from the European Commission through the funding of the ERC Consolidator Grant META-LEGO, N$^\circ$ 101001759.
The authors also gratefully acknowledge the computing time provided on the Linux HPC cluster at Technical University Dortmund (LiDO3), partially funded in the course of the Large-Scale Equipment Initiative by the German Research Foundation (DFG) as project 271512359.
\\
\\
\noindent
\textbf{Conflict of interest.}
The authors declared that they have no conflicts of interest to this work
\endgroup

\begingroup
\setstretch{1}
\setlength\bibitemsep{2pt}
\printbibliography
\endgroup
%
%
%

\end{document}